# Formal Methods:
# From Academia to Industrial Practice
## A Travel Guide

Marieke Huisman

Department of Computer Science (FMT), UT, P.O. Box 217, 7500 AE Enschede, The Netherlands

Dilian Gurov

KTH Royal Institute of Technology, Lindstedtsvägen 3, SE-100 44 Stockholm, Sweden

Alexander Malkis

Department of Informatics (I4), TUM, Boltzmannstr. 3, 85748 Garching, Germany

**Abstract**

For many decades, formal methods are considered to be the way forward to help the software industry to make more reliable and trustworthy software. However, despite this strong belief and many individual success stories, no real change in industrial software development seems to be occurring. In fact, the software industry itself is moving forward rapidly, and the gap between what formal methods can achieve and the daily software-development practice does not appear to be getting smaller (and might even be growing).

In the past, many recommendations have already been made on how to develop formal-methods research in order to close this gap. This paper investigates why the gap nevertheless still exists and provides its own recommendations on what can be done by the formal-methods–research community to bridge it. Our recommendations do not focus on open research questions. In fact, formal-methods tools and techniques are already of high quality and can address many non-trivial problems; we do give some technical recommendations on how tools and techniques can be made more accessible. To a greater extent, we focus on the human aspect: how to achieve impact, how to change the way of thinking of the various stakeholders about this issue, and in particular, as a research community, how to alter our behaviour, and instead of competing, collaborate to address this issue.

# 1 Introduction

Nowadays, software has become an integral part of our daily lives. We can no longer imagine what life would be like if we were not continuously supported by software (and the underlying hardware, of course). As a consequence, there has been an enormous growth in the software industry worldwide, and it is expected that it will continue to increase in the coming years [34, 93]. Moreover, also many other industries such as the service industry (banking, finance [47]) and the automotive industry [83] depend more and more on their software development; they are typically called software-intensive industries and face the same challenges as the software industry.

However, this enormous growth has also made it evident that the software industry is struggling to ensure the reliability of its software [33, 76]. Software failures can have serious economic or societal consequences. For example, recently in Belgium, the ATMs were not usable for a full day after a software update [98], and in the Netherlands, the electronic payment system was intermittently unusable [4, 79]. As a result, banks received damage claims from shop owner organisations, who claimed having a substantial income loss due to software issues. Many other similar examples are available [59, 102], and estimates of the costs of software failures exceed €250,000,000,000 annually worldwide [18, 41, 82]. Moreover, other scientific disciplines also depend increasingly on the reliability of software. For example, a software error recently detected in fMRI software has risked invalidating around 15 to 25 years of brain research including 3500 papers on the subject [29, 30, 71, 72].

For decades, academic researchers have claimed that rigorous use of formal analysis tools can help to increase the quality and reliability of software [22, 104]. A wide range of techniques, with corresponding tool support, have been developed [17]. These techniques differ in the guarantees they provide and in the ease of applicability. There is usually a trade-off: the stronger the guarantees provided by a technique, the more work is typically needed to obtain these guarantees. Despite this variety, all these techniques share a common foundation based on precise mathematical notations (e.g. formal program semantics and program logics), describing the program behaviour and properties [101].

## Success stories of formal methods in industrial practice

Formal analysis techniques have been steadily improving over the last years due to the development of powerful automatic solvers and smart combinations of existing technologies, e.g. in SLAM/SLAM2/SDV [7, 57], Astrée [26, 64], or Frama-C [55]. Multiple examples illustrate that the application of formal methods on industrially-relevant examples is becoming possible. We list some interesting examples here, without striving to be exhaustive.

In the aviation industry the use of formal methods has been integrated in the development standards and accepted as a part of the (mandatory) certification procedure [80, 81]. Tools such as Astrée and Frama-C were successfully employed to formally analyse portions of the code for several aircraft models including the currently largest passenger aircraft A380 [66, 86, 96]. Besides avionics, the Astrée verifier has been routinely applied to the docking software of a cargo space ship, in automotive control, nuclear plant technology, and ventilation [69]. Similarly, in the automotive field formal methods are also gaining increasing attention. Though not strictly enforced by the corresponding automotive standard ISO 26262, some suppliers internally design, check, or verify parts of their software using formal methods [53, 70]; the degree of rigour required by the standard grows along the chain A-B-C-D, from the most relaxed Automotive Safety Integrity Level (ASIL) A to the strictest level D, while formal verification is recommended for C and D [48]. Social networks have no safety-critical software, but they also use formal methods: Facebook internally runs the INFER tool to verify selected properties, such as memory safety errors and other common bugs of their mobile apps, used by over a billion people [21]. The driving force in this case is the huge economic cost of failures. Moreover, in 2011, the AWS division of Amazon started to use TLA+ to meet the requirements stated in their contractual obligations, checking both their present designs and aggressively optimised ones [68]. Amazon believes that formal methods '*accelerate* both time-to-market and quality of [their] projects', and since then, have expanded their efforts, recently also using OpenJML for the analysis of some of their components [25]. Moreover, formal methods have also been successfully used in quite a large number of other areas, for example, to raise the quality of operating system kernels [7, 56], in compilation [60, 65], in telecommunication services [39, 87] to prove or refute properties of cryptography protocols [63], in railway signalling [6, 31], for subway transportation [10, 16], in control systems of the Maeslant storm surge barrier [51, 52, 89] and the Algera bridge [37], for user interfaces [99], in computer-aided design [9], in defence [13], to ensure high quality of cloud services [68], for lighting systems [94], and in a plethora of other areas [5, 27, 77, 78, 104]. Finally, attempts of formal verification of widely used algorithms, protocols, and their implementations sometimes reveal that they are incorrect (e.g. in the case of the Needham-Schroeder protocol [61] or Timsort [28].)

## Formal methods as part of daily industrial practice

Despite the high number of success stories describing the use of formal methods in industrial practice, they did not lead to a systematic integration of formal methods in the daily software-development process [97]. We are not the first to observe this: over the last few decades several papers summarised the state-of-the-art in the use of formal methods in industrial practice and made recommendations on how to strengthen this connection [22, 88, 104]. The recommendations that these papers give regarding research directions, tool integration, etc. have been considered by the industry, but still need further elaboration. Moreover, we see that despite this progress, the software industry changes so quickly that every time academic researchers in formal methods make a step to bridge the gap between formal-methods research and the industrial software-development practice, the gap does not become smaller, because industry has again moved forward, using new technologies and ensuring their market share for the next hype. Furthermore, we think it is important to realise that many of the success stories above depend on an individual academic researcher pushing for it strongly, working hard on building up a relationship with an industrial partner, and adjusting tools to make a specific formal method applicable to the software of this industrial partner.

We believe that to achieve a fundamental change to this situation and to integrate formal methods in the standard software-development process, as a community we have to change the way in which we try to bridge this gap. Instead of individuals pushing forward on their own, we have to act on this together, in a concerted manner. We have to do more than just create individual success stories, but also promote a completely new way of thinking about software development. This paper describes our view on how we, as the formal-methods community, should address this challenge. We will first give an analysis of how we see the current situation, identify what we consider as the bottlenecks, and then present our recommendations. Many of these recommendations are non-technical, but instead, they aim at changing the mentality of the different stakeholders. To structure the analysis and the recommendations, we look at the issue from several perspectives: industry (Section 2), the formal-methods community itself (Section 3), research support structures (Section 4), and education (Section 5). We do realise that our analysis and recommendations may miss out on certain aspects of this complicated issue, but it is our hope that this paper will be a starting point for further discussions on this topic. Ideally, this paper will encourage internal reflection within the formal-methods community, considering how we could become more effective at closing the gap between industrial practice and formal methods.

The analysis and recommendations in this paper are based on numerous discussions we had during workshops and conferences with other researchers and practitioners working in the area of formal methods, who shared their experiences on trying to bridge the gap to industry. We mention, in particular, the discussions held at the workshop 'Verification of Concurrent and Distributed Software', held at the Lorentz Center in Leiden, Netherlands on 14–18 September 2015 [38], and at the track *Formal Methods in Industrial Practice – bridging the gap* that we organised during the 8th International Symposium on Leveraging Applications of Formal Methods, Verification and Validation (ISoLA), 7 November 2018 [32].

## 2 Industry

The term *software industry* is used for a wide range of different companies, ranging from the Googles and Yahoos of this world to single-person companies, developing small mobile phone applications for third-party users. Moreover, many companies in the service industry, such as banks and insurance companies, and even telecommunication companies, are changing into software-intensive companies. This means that the attitude towards software correctness and qual-ity can vary substantially from one software-producing company to another. Furthermore, companies are profit-driven, and to adopt new technologies and processes for quality assurance and software development, they need to back up and justify such changes by a cost-benefit analysis, or at least by perceived pay-offs. Some factors that can influence such decisions are time-to-market, the development and maintenance costs of software, the retail value of the software product (and thus, the added value), the cost of introducing new technology (cost of tools, cost of training engineers or hiring experts, cost of changing the development process), and external requirements, such as what the competitors do in terms of standards and certification.

So, there obviously cannot be a singular approach to quality assurance in the software industry. But the dilemma of how to ensure that software is developed *efficiently*, and at the same time that it functions *correctly* under all circumstances is widely shared within the whole range of software-industry companies. Unfortunately, many software companies do not really perceive that they have such a dilemma. According to Peter Gabriel Neumann, an expert in computing risks and a principal scientist at SRI International, they 'throw this [software] together, shrink wrap it and throw it out there. There's no incentive to do it right, and that's pitiful' [84]. The companies are focused on shipping their software ('time to market') and do not realise that afterwards they spend much longer on maintenance than they would have done on improving software quality during development – provided they had the right tools for it.

Experience with the industrial application of formal methods in certain areas shows that it is indeed beneficial: it results in better code that is easier to maintain and contains fewer errors [73, 74, 92]. However, these applications typically succeed in restricted settings, where the reliability or availability requirements are so high that managers are willing to invest time and effort in adopting formal methods. For the 'mainstream' industry, the situation is different, and the success of these very specific examples is not guaranteed to have an immediate impact on other, less safety-critical or business-critical domains. Moreover, it is very difficult, if not impossible, to come up with concrete numbers for costs and benefits (including later savings) of the application of formal methods.

## Obstacles to the adoption of formal methods

We see several obstacles to the adoption of formal analysis techniques in the software-development process:

- Writing down *software requirements* is often not properly considered as part of the development process; they are left informal, not properly tested, incomplete, etc. However, to use a formal technique, the requirements need to be clarified and formalised, which can only be properly done by programmers with training in requirements engineering. This step incurs fixed costs and takes time: about 30% of the whole time may go into formalisation before any single code line is written, while the remaining phases are shorter [40, Myth 5]. And this step will not be taken, unless there is a perceived immediate pay-off. Moreover, getting the requirements right is often an iterative process, and while developing the system, one needs to go back to the client frequently to validate that the system is still being developed in the desired direction.

- Companies typically employ an abundance of technology (such as build servers, version management, etc.) that supports an established *software-development process*; changes in the development process mean that companies might have to use different technology or adapt the existing one. It is difficult to convince companies to make such a move, because it will incur a production loss (in the short run, at least), and it is difficult for the decision-makers to believe – let alone accept – that the increased software reliability will outweigh this production loss. After all, there are no satisfactory, sufficiently generic studies that compare the effectiveness of various development processes for reaching the same goal. Even for large companies (e.g., BMW) that already employ various software-development processes for different goals, yet another, new process would require education and a change of mindset, a change that incurs resistance in any case.

  In this context, it is important to note that the adoption of formal technologies can vary significantly in the impact it has on the software-development process. Verification is less disruptive to established processes, but there is a trade-off between the effectiveness of the verification and system size: early verification can effectively ascertain that a specification is correctly represented by its slightly larger refinement, whereas post-design verification is less likely to be effective, since it has to deal with a large system already built. For instance, early verification is the contents of the Design Verification phase of the V-Modell XT [3, G.1.1.5], while post-design verification can be used (and is sometimes used) to reduce the testing burden in classical software-development processes following a Top-Down approach [91, Abb. 1.1]. Model-driven design where specification and verification are integrated into an agile development process is more promising, since it is incremental, but requires a new process and may be more difficult to adopt. For example, (formal) specification can be an item on the backlog and a subsequent goal of a Scrum [103] sprint, and verification can be used in the Testing phase of Design Thinking [100] if the properties to be verified are clear. Correctness-by-construction is another promising paradigm, but also requires a substantial change in the way systems are designed. Synthesis is a perfectly valid backlog item in Scrum, of course, assuming that a formal specification exists and the product owner agrees to it.

- Software developers (or their managers) are often simply unaware of the *wide range of tools and techniques* available that could help them to improve their software quality. And even if they are aware, there is never any time to quietly search for available options or to try out some unknown technology because of the enormous market pressure. Moreover, unfortunately, it is very hard to make a clear and indisputable estimate of the expected cost reduction.

- Existing tools and technologies *never fit one-on-one*. Nowadays, software developers often use very complex development environments, which make extensive use of different frameworks and libraries (which are often developed in different programming languages and can be frequently changing). Time is needed to adapt to the format as required by the formal method at hand. This time is frequently not available, because of the time-to-market pressure, and even when it is, there are always other, more pressing matters that the decision-makers feel the need to address first. Furthermore, the overwhelming majority of the formal-methods tools are typical research prototypes: they do not work under all circumstances, do not cover all cases, and are not professionally maintained. And if they do work well and are maintained, they are often very expensive – probably because of the small market for them. However, in general, formal methods need not be intrinsically costly; they do pay off when applied properly: 'The fact is that writing a formal specification *decreases* the cost of development' [40, Myth 5].

- Then, *human psychology* also comes into play: programmers are reluctant to change the way they have always programmed, and managers do not want to adopt novelties they do not understand.

- Finally, we should keep in mind that many programmers have not taken any *appropriate courses* at research universities. In the USA, for instance, 35% of computer programmers had no Bachelor's or higher degree as of year 2002 [90, p. 98]; according to a worldwide online survey [46], about one fourth of professional developers had no Master's or higher degree in 2019. Such programmers might have never completed any formal education at all, or they might have attended high school only, or they might have completed some level of professional, non-research education, e.g. a degree at a university of applied sciences[1], etc. At the same time, many programmers that do have a higher education are not necessarily trained in Computer Science.

## Enablers for the adoption of formal methods in industry

We should first look more seriously at the current (informal or semi-formal) validation methods being used in industry, such as testing, code inspection, etc. Methods and processes for software testing are being rapidly developed, and software-testing research is a hot topic within software engineering. Testing is now also being increasingly automated and integrated in agile software-development processes with frequent builds. These methods and processes are a natural place to gradually introduce formal methods in industry by successively adding automated tool support in the testing and verification process instead of selling formal methods to industry as a 'standalone' technique. Another interesting option is to try to combine formal methods and testing to make the full verification process more efficient, e.g. by using formal methods to direct testing efforts to the 'dark corners' typically discovered by such methods.

[1]As can be found in Germany and the Netherlands, for example.

At the same time we feel that eventually something more is needed, namely that *the application of formal methods becomes an integral part of the software-development process*. This can start with small enhancements, such as extending the common usage of static type checking (e.g. automatic warnings if a null-pointer exception or some other run-time error could occur), and should finally lead to full and seamless integration of formal methods in the software-development process. Also tool-combining platforms, which make the reuse of results possible and enable the use of machinery from common development platforms such as Eclipse, are important first steps in this direction [62], but more effort is needed for a smooth integration into the development process.

Then, it is also important to involve the *application expert* directly in the development so that they feel the impact of formal methods themselves [35, 67]. The point here is to find a match to the application expert's mindset. We cannot expect to educate everybody to become a formal-methods expert; instead, it should be up to us to try to get closer to the customers, domain experts, and users by really trying to understand their way of thinking and getting them to use formal methods.

Eventually, what we as a formal-methods community should continue to aim for is that the people involved in the software industry will become aware that formal methods are useful and will consider the application of formal techniques as a first option to ensure software quality rather than as a last resort. This is a change in attitude taking many years, which can only be achieved by tackling the issue from many different angles, not by simply telling industrial software developers that formal methods will solve all their problems.

## Recommendations

1. Invest time in industrially-relevant case studies in order to understand what techniques are actually needed for industrially-relevant applications. Study existing tools and practices and investigate how to integrate frequently used libraries and frameworks in formal verification.

2. Investigate and understand (semi-formal) processes and methods currently used in industry and see how these can be further improved by combining them with formal methods.

3. When designing a formal-methods–supported software-development process, keep the implementers in mind and adapt social deterrents and incentives correspondingly, so that the implementers are motivated to continue to use those techniques.

4. Train engineers and managers in the software industry in good software development and seamlessly integrate formal methods in the process, so that in the long run the engineers and managers will see this as simply a part of the process, not as a special add-on.

5. Investigate simple and lightweight ways to integrate the use of formal methods in the software-development process, ensuring that if a developer tries a formal method for the first time, it will be a pleasant experience, with most likely a high return on investment.

6. Identify areas where formal methods in an industrial setting are most useful and are most easily applied and make sure these are shared among the formal-methods community in order to create a common awareness of good targets.

7. Continuously try to convince your industrial partners of the benefits and opportunities offered from applying formal methods and from hiring staff that has training in such methods, since only in that way the methods will eventually sink in.

8. Share your success stories of applying formal methods in an industrial setting.

# 3 Formal-methods community

Next, we consider the formal-methods community itself and what we, as academic researchers, can do to bridge the aforementioned gap between formal methods in academia and daily industrial practice. Some of the issues that we identify are related to academia in general, whereas some are more specific to the formal-methods community.

## General challenges for academia

An important challenge we have to face here is that academic success is measured by publications, projects, and prizes. This encourages individualism: collaboration and putting effort into making something usable for others does not necessarily advance one's own career. These are the 'rules of the game' that many people play without questioning. As a consequence, we all like our own approach best, and we do not appreciate each other's work, especially if it is not sufficiently 'new'. And if somebody does something that is new, we are mainly interested in how we can beat this person and come up with something that is even better and newer. But is this the best way for society to advance science? As a consequence of the current set-up, academic researchers focus on solving the most difficult and challenging problems and on how to find intricate solutions for those. We then develop a solution that works for this particular challenging aspect of the problem; but we have typically no reason or incentive to adapt this to a complete solution that is applicable in the 'real world'.

What should be done about this? We believe that the existing set-up is still important for us as academic researchers that keep on looking for scientifically challenging problems, and we should keep on striving to find the most elegant solutions; after all, this is what we are good at, and this is our added value. **However**, we should also look at how we can join forces, collaborate, and in this way, produce new results and enlarge the scale of what we can do. We should also be open for the problems that industry faces and help industry solve them by finding the best solution at hand as opposed to simply our own solution. This will require a change of mindset of many researchers, as they need to realise the added value of collaboration, e.g. potentially higher impact. Such a change will also require new kinds of incentives from our academic institutions and the funding structures.

## Specific challenges for the formal-methods community

To make this change of mindset possible, we as a community should appreciate the effort that is put into *tool development* more. There are already some conferences that explicitly encourage this (e.g. FMICS, ISoLA, iFM, TACAS, FASE), but we feel that more conferences should have special tool paper categories, where the effort (typically illustrated by an industrial case study) of making a tool widely usable or of integrating different techniques is appreciated. Of particular importance are also journals which are dedicated to the aspect of technology transfer by means of tools and case studies. STTT [23] and *Stories from the Front* in the Journal of Systems and Software [85] are good examples of this, but we think more journals should encourage such submissions. In this way, even if we do not change the overall system of appreciation, we can still make sure that researchers can obtain credits for tool work. Teaming up with software- and requirements-engineering researchers can also be an important way to achieve this: for instance, they have established venues where industrial case studies are welcome and often also have many industrial contacts. Another way to give scientific credits to work on tool development is to organise competitions. There are already several such competitions: VSComp [24], the VerifyThis program-verification competition [44, 45], SV-Comp [11], the RERS competition [49], the SAT competition [8], etc. (see [12] for a more complete overview). A good result in such a competition is considered as a scientific achievement. Moreover, competitions also make tool developers more aware of what can be achieved with other tools, so that they can learn from each other, discuss, and exchange efforts.

Additionally, we as a community should put effort into advertising our tools, e.g. by making YouTube movies, and into providing online platforms to use the tools. It is important that we exchange ideas and help each other to achieve this. We should also make a joint effort to advertise ourselves on the Web (Wikipedia, Facebook, etc.).

### Recommendations

1. As a community, collaborate on advertising our tools and how they could be used together in order to create maximal visibility of what we can achieve for the software industry.

2. Put effort in changing the mindset of what is necessary to be a good researcher and make sure that collaboration is encouraged to create the necessary space and time for researchers that work on industrially relevant activities.

3. Make sure sufficient scientific credits can be obtained for the effort put into tool development, industrial case studies, and technology transfer by having more conferences and journals that report on such activities and by teaming up with requirements- and software-engineering researchers.

## 4 Research support

A major challenge that we see is how to obtain (financial) support for research activities that are aiming at bridging the gap between industry and academia. To achieve this, researchers first of all need to invest time and effort in building up a relationship with potentially interested industrial partners, e.g. by performing a case study to solve some problem the company is interested in. Only once this relationship has been established will companies be open to listen to and understand the innovations proposed by researchers. However, finding time is difficult, since searching for industrial partners is only one of the many activities that researchers have to carry out. Moreover, industrial case studies can be difficult to publish, for they seldom describe new ideas, but tend to apply existing techniques and tools in a particular setting.

Researchers typically get funded to develop new ideas and to publish papers describing these new ideas. For research papers, it is sufficient to develop a prototype, i.e., a proof-of-concept tool that can demonstrate that the ideas work and can be implemented (typically at the *Technology Readiness Level*s (*TRL*s) [36] 3 to 5). However, such a prototype tool typically cannot be directly transferred to industry. For a tool to be usable in an industrial setting, it needs to be robust and provide full coverage (i.e. at least at TRL 8). For example, for a prototype program-analysis tool it can be sufficient to completely ignore exceptional control flow, but when such a tool is used in a realistic software-development setting, it also needs to handle this aspect.

Extending a tool to make it usable and robust in an industrial setting typically requires much engineering work, for which often a programmer or engineer is better suited than a researcher. In the academic world, however, no credits are normally given for such activities, and *limited funding is available*. If a researcher wishes to undertake this kind of work, usually the only way to do this is to start a spin-off company. However, not all researchers are interested in setting up companies, and moreover, creating a successful company requires a completely different skill set.

We believe that if funding agencies (and governments) consider that technology transfer from academia to industry is important, they should support this by dedicated funding schemes, which can support work on tool development for a certain period. Importantly, such a funding scheme should not require that the outcome is a tool that can be directly commercialised; the goal should be to develop a tool that is demonstrable to end-users (typically TRL 6–7). It might still lack functionality in some corner cases, but these corner cases should be clearly defined, and there should be a substantial class of applications for which the tool just works without complications. We believe that it is necessary to reach such a state before a company could be convinced to invest time and money in the further development of such a tool.

The development of this end-user–demonstrable tool does not necessarily have to be carried out at the university. For example, many different European countries have created a national 'eScience center' [1], which develops software and methods for the scientific community; such an institution can also be a perfect place to extend prototype tools into something that can be demonstrated to industry.

### Recommendations

The following recommendations refer to researchers that are in the position to influence the policy makers of the research funding agencies.

1. Develop flexible funding schemes (in-cash or in-kind) to support the engineering work that is necessary to transform a prototype implementation into a demonstrable implementation.

2. Make sure that also researchers that do not want to create a company feel an incentive to transfer their techniques to industry to ensure that the most promising ideas will actually be developed further into prototypes, independent of the affinity with commercial activities of the researcher.

3. Ensure that academic credits can be obtained by transferring results to industry, for example, by giving more priority and weight to conference tracks and journals soliciting tool and demo papers (and making sure that these papers are indeed reviewed as tool-and-demo papers, not as regular research papers). Put emphasis on and reward such activities when recruiting new or promoting existing staff, such that researchers in the formal-methods community feel that activities in this direction will not have a negative impact on their scientific ambitions.

## 5 Education

Currently, in university education programmes, formal methods are typically taught in a separate course – and often have a reputation of being difficult.

One of the reasons for this is that we as teachers like to teach our most recent developments, giving difficult and challenging verification tasks to the students. We should **not** stop doing this, but we should realise that this only attracts a small percentage of students, and if we want to have impact, we need to broaden the target group of formal-methods–related education.

In particular, we should put additional effort into familiarising **also** the larger percentage of the students with formal techniques and what can be achieved by applying them. Therefore, *formal methods should be woven into software-development and requirements-engineering courses.* This should target not only software-development courses taught at universities, but also software-development courses at other levels (vocational education, training for software developers, etc.). In these courses, we should not force all students to prove full correctness of whatever they develop. Instead, we should give them a good feeling of what can be achieved by using more rigorous and formal techniques to support their requirements-engineering and software-development processes and a good feeling for the wide range of tools and techniques that are available. In particular, they should understand that rigour comes at a cost: the effort to achieve that. It is important that in these courses, lecturers encourage the students to use the tools and techniques that are reasonably well-developed and stable and that will convince the students that the usage of these tools really helps to improve software quality (i.e. they should experience the reduction in development and maintenance time). This means that lecturers should not always use their 'own' tool in such courses, but they should put real effort in using tools and techniques that are adequate for the job at hand. It is important that these techniques are taught as something that is simply *part of the process* and not as an optional add-on.

As a concrete case of how this can be achieved, one of the authors has introduced the writing of JML specifications to document the code as part of the first programming course in Java [42]. This forces the students to think about the behaviour of their code and write this in a formal way. In addition, the course provides extra exercises and an optional lecture where the students are challenged to use the run-time checker to validate their specifications. In later years of the programme, this run-time checker can be used as a mandatory step in software development (and static verification is presented as an option). Full verification is taught in a Master's course, which is mandatory for students in software technology and embedded systems [43]. This course does not focus on the details of how the formal methods are implemented, but rather concentrates on what actually can be achieved with formal techniques. Courses with similar ideas are offered by other universities; concrete examples are [2, 15, 19, 20, 58]. In the literature, this approach to integrate formal methods in the regular software-engineering training has been coined *'Secret Ninja Formal Methods'* [54].

Teaching the use of formal methods as an integral part of the software-development process requires that this idea is rolled out throughout the whole educational programme. Moreover, it also requires that there are some high-quality tools robust enough to be used by students.

An important means to achieve this state of affairs is by exchanging teaching experiences and best practices within the formal-methods community. There are Web pages like Formal Methods Europe [50] and workshops on teaching formal methods such as [14, 75]. Such venues should be much more actively used to distribute these ideas (and provide information for reuse).

Moreover, we should in addition keep our high-level expert formal-methods teaching for the people that really want to understand how things work and, in this way, ensure that the research in formal methods keeps on going strong. For these specialist courses, we should consider how to enlarge the audience by enabling students from other universities to follow such courses as well. One possibility for this is to develop a Massive Open Online Course [95] on such a specialised topic.

Finally, we should keep in mind that only focussing on academic education is not sufficient to make all software developers familiarise themselves with formal methods. As mentioned above, when discussing the situation in industry, many software developers are not trained as academic computer scientists, but learn programming in some other way (e.g. at high school, at a university of applied sciences, or as part of a post-doctoral education). While we do not have the means to educate all these programmers directly in proper software and system development, we should make an effort to educate their teachers and aim for a transfer effect.

## Recommendations

1. Teach both a specialised formal-methods course and a course where formal methods are simply part of the process – without emphasising that the second course teaches formal methods – so that future software developers see the use of formal methods as an integral part of the requirements-engineering and software-development process and experience the benefits of formal methods already during training.

2. Ascertain that students get a positive experience working with formal tools and techniques, so that they see the benefits and do not immediately, collectively decide that this is not useful.

3. Do not insist on using only your own tools and techniques, but use well-established and stable tools, again to ensure that students have a positive experience, and do not consider the application of formal methods to be a time-consuming struggle.

4. Make specialised courses available to a large audience to have a maximal dissemination of this specialist knowledge.

5. Offer courses for school teachers and teachers in vocational education and provide them with suitable training materials for their students, so that they can pass on their experiences to more mixed groups of students.

# 6 Conclusions

In this paper, we have considered the gap between academic research and industrial practice in the area of formal methods. We have looked at this issue from different perspectives: industry, education, research support, and the formal-methods community itself. We believe that from all these points of view, the formal-methods community should be able to take steps towards closing this gap, and we have provided a concrete list of recommendations.

Our conclusions are drawn from a number of *observations*, which can be summarised as follows:

1. Computer Science is a rather new and immature discipline, which, together with certain historical factors, has lead to a mutual distrust between industry and academia in that field and to a reluctance to collaborate.

2. Companies are profit-driven, while academic researchers are novelty-driven; this discrepancy has only increased the gap between the two worlds.

3. There are a number of success stories of applying formal methods in industry, but these have largely been the result of individual efforts and not of consorted effort incentivised by funding agencies.

4. At the same time, there are ample opportunities for technology transfer of formal methods to industry that have not been utilised.

Our *recommendations* to the formal-methods community can be summarised as follows:

1. We need to learn how industry works and be willing to do industrially relevant research and industrially useful work.

2. Universities need to find ways to incentivise industrial collaboration by adjusting its system of academic and career credits.

3. The research support and funding agencies need to actively encourage tool development and maintenance beyond prototyping.

4. In academic education, we should ensure that formal tools and techniques become an integral part of software-development teaching; where appropriate, we should also teach how to conduct industrially relevant research.

What we have seen is that there are already some industrial areas that have an interest in software verification, in particular, in the safety-critical software industry. This is good, and we should definitely continue to collaborate with them and have impact there. Furthermore, we should use the successes in these areas as a lever to reach out to other, less safety-critical areas of software development. Only if in the end we manage to convey our theoretical results in formal methods into the daily software-development process of all those small and medium-sized enterprises out there can we conclude that we have finally closed the gap between academia and industry. To reach this goal, the formal-methods community should change its mentality, actively collaborate, and encourage each other to achieve maximal industrial impact.

# Acknowledgements

We are indebted to Christian Prehofer, Bernhard Steffen, and Björn Lisper for giving useful feedback on earlier drafts of this paper. In particular, Björn Lisper pointed out the importance of investigating the techniques currently used in industry and the advantages of teaming up with software-engineering research, while Bernhard Steffen pointed out many relevant initiatives that are working towards the same goal as we intend with our paper. Bernhard Steffen also emphasised the importance of *lightweight* formal methods that make it possible to involve the application expert directly. Finally, Christian Prehofer pointed out that it is impossible to precisely define the costs and benefits of formal methods as well as the importance of taking into account which frameworks and libraries are frequently used in industry and incorporating them in the various academic formal-methods prototypes. We are further indebted to Elizabeth Hamzi-Schmidt for proofreading parts of the paper.

Funding: This work was supported by the NWO 639.023.710 VICI project Mercedes; the ITEA-3 project REVaMP2 [project number 15010]; and the Software- and Systems-Engineering Research Group at the Technical University of Munich, Germany.